\documentclass[onecolumn,showpacs,preprintnumbers,amsmath,amssymb]
{revtex4}
\usepackage{subfigure}
\usepackage{graphicx,dcolumn,bm,color,latexsym,amssymb}
\usepackage[latin1]{inputenc}
\usepackage{import}
\usepackage{xifthen}
\usepackage{pdfpages}
\usepackage{transparent}
\usepackage{verbatim}

\definecolor{Blue}{rgb}{0.0,0.0,1}
\definecolor{Red}{rgb}{1,0.0,0.0}
\definecolor{Green}{rgb}{0,0.5,0.0}

\begin{document}

\title{Heat fluctuations in the logarithm-harmonic potential} 

\author{Pedro V. Paraguass\'{u}}
 \email{paraguassu@aluno.puc-rio.br}
\affiliation{Departamento de F\'{i}sica, Pontif\'{i}cia Universidade Cat\'{o}lica\\ 22452-970, Rio de Janeiro, Brazil}

\author{Welles A.~M. Morgado}
 \email{welles@puc-rio.br}
\affiliation{Departamento de F\'{i}sica, Pontif\'{i}cia Universidade Cat\'{o}lica\\ 22452-970, Rio de Janeiro, Brazil\\ and National  Institute of Science and Technology for Complex Systems}

\date{\today}

\begin{abstract}
 Thermodynamic quantities, like heat, entropy, or work, are random variables, in stochastic systems. Here, we investigate the statistics of the heat exchanged by a Brownian particle subjected to a logarithm-harmonic potential.  We derive analytically the characteristic function and its moments for the heat. Through numerical integration, and numerical simulation, we calculate the probability distribution as well, characterizing fully the statistical behavior of the heat. The results are also investigated in the asymptotic limit, where we encounter the characteristic function in terms of hypergeometric functions. 
\end{abstract}

%\pacs{???}

\maketitle

\noindent Keywords: heat fluctuations; stochastic thermodynamics; exact results.
%\maketitle

%\pacs{05.40.-a, 05.10.Gg, 05.70.Ln}

%
%	1	1	1
%
\section{Introduction}

Diffusive systems such as Brownian particles exist in mesoscopic scales, ranging from a few nano-meters to micro-meters, where thermal fluctuations play an important role. In  general,  these  systems  are  far away  from  equilibrium,  and  their  thermodynamic quantities,  like  heat, or  work,  become  random  variables. In  order  to  describe  the thermodynamics  of  mesoscopic  systems,  we  can  use  the  Stochastic  Thermodynamic framework \cite{oliveira2020classical,ciliberto_experiments_2017,seifert2012stochastic,sekimoto2010stochastic}. In this framework, thermodynamic quantities become stochastic, having an associated probability distribution. This can be contrasted to the usual thermodynamic quantities which have negligible fluctuations due to thermodynamic limit.

Characterization of the statistics of the thermodynamic quantities of diffusive systems was carried in many different models. Derivations of the distribution and characteristic function for work and heat had been solved for Brownian systems, ranging from the free particle to non-harmonic potentials with theoretical \cite{paraguassu_heat_2021,chatterjee_exact_2010,chatterjee_single-molecule_2011,ghosal_distribution_2016,saha_work_2014,kusmierz_heat_2014,chvosta_statistics_2020,goswami_heat_2019,goswami_work_2019,goswami_work_2021,kwon_work_2013,jimenez-aquino_thermodynamic_2018} and experimental \cite{joubaud_fluctuation_2007,gomez-solano_heat_2011,imparato_work_2007,imparato_probability_2008} results. These works bring physical insights in the thermodynamics of such systems. In particular, we are interested in investigating the fluctuations of heat in a diffusive system with logarithm-harmonic potential. The fluctuations can be understood by the probability distribution or the characteristic function. The later is more easily derived, however, the distribution is also desired, since allows us a direct interpretation of its behavior through a more simple visualization. The physical picture we have in mind is that of an electrolyte interacting with a charged polymer by means of an electric interaction and a harmonic potential.

Concerning the logarithm potential, the heat distribution for the case without the harmonic term was already studied by the authors in \cite{paraguassu_heat_2021}. Moreover, the work for in the logarithm harmonic system was studied in \cite{ryabov_work_2013,holubec_asymptotics_2015,ryabov2015stochastic}, where the time-variation of the stiffness of the harmonic term allows us to extract work from the particle. While the work properties was already investigated in \cite{ryabov_work_2013,holubec_asymptotics_2015,ryabov2015stochastic}, the heat distribution is still not studied, asfar as we know. Hence, we present herein a comprehensible investigation of the heat distribution, obtaining its probability and characteristic function. 

Diffusive systems under a Logarithm potential have stirred some interest in the literature \cite{leibovich_aging_2016,hirschberg_approach_2011,aghion_non-normalizable_2019,dechant_superaging_2012,barkai_area_2014,kessler_infinite_2010,ray_diffusion_2020} due to its intriguing effects, such as ergodicity breaking \cite{dechant_superaging_2012}, anomalous diffusion~\cite{barkai_area_2014,kessler_infinite_2010}, and resetting phenomena \cite{ray_diffusion_2020}, to name but a few. Moreover, the logarithm potential can represent different systems \cite{bray_random_2000,ryabov_brownian_2015,fogedby_dna_2007,lo_dynamics_2002,campisi_logarithmic_2012,kessler_infinite_2010,barkai_area_2014}. Regarding the Brownian particle, the logarithm potential can represent the Coulomb force of a line of charges \cite{manning_limiting_1969,liboff_brownian_1966}, which is used to model a long polymer interacting with a charged Brownian particle \cite{manning_limiting_1969}. In our case, in addition we have a harmonic potential, which can represent an asymmetric trap potential \cite{ryabov2015stochastic} allowing us to have a defined equilibrium distribution.

In stochastic thermodynamics, heat is in a sense a more fundamental quantity than work. Since, to exchange work with an external system, it is necessary that the system  interact with the external system in an ordered way. Conversely, the heat will always be present since it is the energy exchanged between the particle and the heat bath \cite{sekimoto2010stochastic} in a disordered, and spontaneous, way. Nevertheless, because of its definition, heat cannot be measured directly for a Brownian system. An alternative is to measure the trajectories and then construct the statistics of the heat \cite{ciliberto_experiments_2017}. Therefore, a theoretical investigation is always desired to compare with experimental results. Theoretical heat distributions results can be found in \cite{paraguassu_heat_2021_2,paraguassu_heat_2021,gupta_heat_2021,fogedby_heat_2020,goswami_heat_2019,crisanti_heat_2017,ghosal_distribution_2016,rosinberg_heat_2016,kim_heat_2014,kusmierz_heat_2014,saha_work_2014,chatterjee_single-molecule_2011,chatterjee_exact_2010,fogedby_heat_2009,imparato_probability_2008,imparato_work_2007,joubaud_fluctuation_2007}. Here, combining analytical methods with numerical simulations  we derived the heat distribution for the logarithm-harmonic potential, extending the list of results in heat distributions.

The present paper is organized as follows: In section 2 we describe the dynamics and thermodynamics of the Brownian particle in a logarithm-harmonic potential. In section 3 we derive the probability distribution of the heat. In section 4 we analyze the characteristic function and calculate the moments of the distribution. In section 5 we give the details of the numeric simulation. We finish in section 6 with the discussion of the results and the conclusion.

\section{Stochastic Thermodynamics and Dynamics of the particle}\label{sec2}

In this paper, we consider a diffusive particle, in the overdamped limit, under a logarithm and a harmonic potential. The logarithm potential can represent the potential of a line of charge, while the harmonic potential is often used to model the action of optical tweezers. Together, the potentials can represent an asymmetric trap. Here, following \cite{bray_random_2000,ryabov_work_2013,giampaoli_exact_1999} we consider that the position $x(t)$ is only defined in the positive values of the $x-$axis. With this configuration, we have the Langevin equation:
\begin{equation}
    \dot{x}-\frac{g}{x}+kx = \eta(t),\label{langevin}
\end{equation}
where the force $-g/x$ comes from the logarithm potential, while the force $kx$ comes from the harmonic potential. To model a heat bath, the noise is defined as a Gaussian white-noise, that is
\begin{equation}
    \langle \eta(t)\rangle =0,\;\;\; \langle \eta(t)\eta(t') \rangle = 2T \delta(t-t'). 
\end{equation}
Here we are working in the units where $\gamma, k_B =1$. Then, only the temperature $T$ appears in the correlation.  Here, we shall call it the strength of the heat bath, while the other parameters, $k$ and $g$, are the strengths of the internal potential. 

\begin{figure}
    \centering
    \includegraphics[width=17cm]{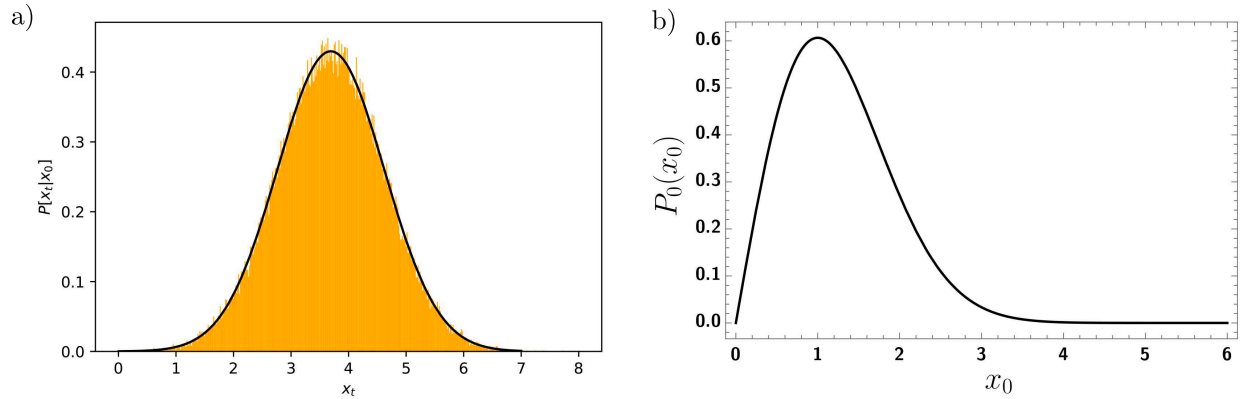}
    \caption{a) Conditional probability with $x_0=10, T,k=1, g=0.1$ and its numerical simulation of histogram in orange. b) Normalized canonical initial distribution with $g,T,k=1$.}
    \label{fig1}
\end{figure}

For the case where the Coulomb force is attractive, the system exhibits the first passage problem as studied in \cite{ryabov_brownian_2015}. However, here we will focus on the case where the potential is strong repulsive, where it is possible to have an equilibrium initial distribution for the position, that is
\begin{equation}
     P_0(x_0) = \frac{2^{\frac{1}{2} \left(1-\frac{g}{T}\right)} \left(\frac{k}{T}\right)^{\frac{g+T}{2 T}} }{\Gamma \left(\frac{g+T}{2 T}\right)}\exp\left(\frac{g \log (x_0)}{T}-\frac{k x_0^2}{2T}\right)\label{initial}
\end{equation}
 where the relation
\[
    g\geq T,
\] 
guarantees the convergence of the initial distribution and yields the strong repulsive behavior. Moreover, with this constraint,  the potential $V(x)=\frac{k}{2}x^2-g\log x$ resembles  an asymmetric trap \cite{ryabov_brownian_2015}. The initial distribution is plotted in figure \ref{fig1} b).

The conditional probability is an important quantity in the derivation of the heat distribution. For the logarithm-harmonic case, it could be derived by path integrals \cite{giampaoli_exact_1999} or via the Fokker-Planck equation \cite{ryabov2015stochastic,dechant_solution_2011}. The normalized conditional probability is given by : 
\begin{equation}
  P\left[x_t, t | x_{0},0\right]=\frac{x_{0} \mathrm{e}^{(\nu+2)  kt }}{2 \frac{T}{k}e^{kt}\sinh{kt}}\left(\frac{x}{x_{0}}\right)^{\nu+1} \exp \left(-\frac{x_t^{2} \mathrm{e}^{2 kt }+x_{0}^{2}}{4 \frac{T}{k}e^{kt}\sinh{kt}}\right) \mathrm{I}_{\nu}\left(\frac{x_t x_{0} \mathrm{e}^{ kt }}{2 \frac{T}{k}e^{kt}\sinh{kt}}\right),
\end{equation}
which is the same conditional probability derived in \cite{ryabov_brownian_2015}, but here we work with $k$ as a constant. This distribution is plotted in  figure \ref{fig2} a). We also the results from a simulation of this distribution, which is a sort of preview for the simulation of the heat distribution.

Since there is not any changes of  parameters of the system, there is no  work made on the system. Consequently, the stochastic thermodynamics of such a system is characterized solely by the heat. Following Sekimoto  \cite{sekimoto2010stochastic}, the heat is defined as the energy exchanged between the particle and the heat bath, which is
\begin{equation}
    Q[x(t)]=\int_0^t \left(-\dot x +\eta(t)\right) dx
\end{equation}
where the integral is defined in the Stratonovich prescription, which is an appropriate prescription to work with in stochastic thermodynamics \cite{bo_functionals_2019}. Using the Langevin equation \ref{langevin} we derive the more simple formula
\begin{equation}
    Q[x(t)] = \frac{k}{2}(x_t^2-x_0^2) - g\log{\left(\frac{x_t}{x_0}\right)}\label{heat}
\end{equation}
where one can notice that the heat is just the difference in the internal energy $\Delta U=Q[x]$, besides  the fact that the heat can be written in terms of the final and initial points of the trajectory. We will see in the next section that the heat exhibits a  non-trivial statistical behavior.

\section{Derivation of the Heat distribution}\label{sec3}

The heat, being a functional of the trajectory, has the conditional probability
\begin{equation}
    P(Q|Q=Q[x])=\delta(Q-Q[x]).\label{cond}
\end{equation}
It emphasizes that the random values of $Q$ are given by the trajectory-dependent formula $Q[x]$. In the studied case, the heat only depends on the initial and final points of the trajectory. Therefore, the probability distribution for the heat will be given by
\begin{equation}
    P(Q)=\int dx_t dx_0 P_0(x_0) P[x_t,t|x_0,0]\delta(Q-Q[x]).\label{pq}
\end{equation}
Moreover, we can rewrite  \ref{pq} as
\begin{equation}
    P(Q)=\int \frac{d\lambda}{2\pi} e^{i\lambda Q} Z(\lambda),
\end{equation}
where
\begin{equation}
    Z(\lambda) = \int dx_t\int dx_0 P_0(x_0) P[x_t,t|x_0,0] \exp(-i\lambda Q[x(t)]) .
\end{equation}
In the above formula, we just write the Dirac delta as an integral over $\lambda$. This formula is useful since allows us to identify $Z(\lambda)$ as the characteristic function.

The integrals in $Z(\lambda)$ can be carried analytically. But, the calculations are long, and the manipulations can be simplified using Mathematica \cite{Mathematica,abramowitz1988handbook}. The important thing to notice is the following integral identities:
\begin{equation}
  \int_0^\infty I_\nu(ax)x^be^{-cx^2} dx = 2^{-\nu -1} a^{\nu } c^{\frac{1}{2} (-b-\nu -1)} \Gamma \left(\frac{1}{2} (b+\nu +1)\right) \, _1\tilde{F}_1\left(\frac{1}{2} (b+\nu +1);\nu +1;\frac{a^2}{4 c}\right), \label{id1}
\end{equation}
where $_1\tilde{F}_1$ is the regularized Kummer confluent hypergeometric function \cite{abramowitz1988handbook}, and the conditions $a>0,b>-1,c>0$ have to be satisfied. And 
\begin{equation}
     \int_0^\infty {}_1\tilde{F}_1\left(a;b;cx^2\right)x^de^{-fx^2}=\frac{1}{2} f^{-\frac{d}{2}-\frac{1}{2}} \Gamma \left(\frac{d+1}{2}\right) \, {}_2\tilde{F}_1\left(a,\frac{d+1}{2};b;\frac{c}{f}\right),\label{id2}
\end{equation}
where $_2\tilde{F}_1$ is the regularized Gaussian hypergeometric function \cite{abramowitz1988handbook}, and $d>-1$. The identities are used to integrate over $x_t$ and $x_0$. We use Eq.~\ref{id1} to integrate over $x_t$ and Eq.~\ref{id2} to integrate over $x_0$. 

\begin{figure}
    \centering
    \includegraphics[width=16cm]{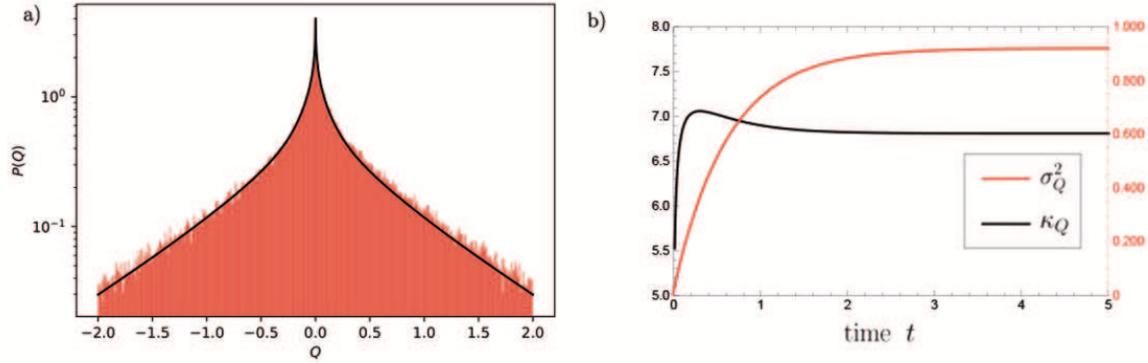}
    \caption{a) Heat distribution in logarithm scale with simulation values. b) Plot of the variance and the kurtosis. Variance (red line) with the vertical axis on the right (red axis) and kurtosis (black line) with the vertical axis on the left (black axis). We only have kurtosis and variance, since the mean and skewness are zero. The parameters values are $g=0.1, k=1, T=1,t=1$.  }
    \label{fig2}
\end{figure}

After integrating over $x_t$ and $x_0$ we find $Z(\lambda)$, which is written explicit in the next section (see Eq.~\ref{char}). With $Z(\lambda)$ we still need to integrate over $\lambda$ to find the probability distribution. This integration can only be carried numerically, and the result is plotted in figure \ref{fig2} a).

\section{Characteristic Function and Moments}\label{sec4}

Even though the probability density needs to be calculated numerically, the characteristic function is achieved naturally as an intermediary step in the derivation of the probability density. The characteristic function is then
\begin{eqnarray}
 \nonumber    Z(\lambda)=  \frac{(2k)^{\frac{g+T}{T}}}{\Gamma \left(\frac{g+T}{2 T}\right)^2}  \left(2\sinh{kt}\right)^{-\frac{g+T}{2 T}}  f(\lambda)^{-z_\lambda} f({-\lambda})^{-z_\lambda-ig\lambda} \Gamma \left(z_\lambda\right) \Gamma \left(z_\lambda+ig\lambda\right) \times \\  _2F_1\left(z_\lambda,z_\lambda+ig\lambda;\frac{g+T}{2 T};\frac{k^2 \text{csch}^2(k t)}{\coth (k t) (\coth (k t)+2) k^2+k^2+4 T^2 \lambda ^2}\right),\label{char}
\end{eqnarray}
where we defined the functions
\begin{equation}
    f(\lambda) = (k-2iT\lambda+k \coth{kt}),
\end{equation}
\begin{equation}
    z_\lambda = \frac{g+T-igT\lambda}{2T},
\end{equation}
for shortness sake, and the function $_2F_1(...)$ is the Gaussian hypergeometric function \cite{abramowitz1988handbook}.
With the above formula, we can simply check that the normalization of the probability is satisfied
\begin{equation}
    Z(0)=1.
\end{equation}

\subsection{Moments}
With the characteristic function  \ref{char}, we can calculate all the moments of the distribution, characterizing the fluctuations of the heat. The moments are given by
\begin{equation}
    \langle Q^n \rangle = (-i)^n \frac{\partial^nZ(\lambda)}{\partial \lambda^n}\mid_{\lambda=0}.\label{average}
\end{equation}
Then, using the above formula, we can calculate the mean and the second moment:
\begin{equation}
    \langle Q\rangle = -i \frac{\partial Z}{\partial \lambda}\mid_{\lambda=0} = 0,
\end{equation}
\begin{eqnarray}
     \langle Q^2\rangle = \frac{1}{4}\left(g^2 \mathcal{J}(t) \sqrt{1-e^{-2 k t}} e^{-\frac{g k t}{T}} \left(e^{2 k t}-1\right)^{\frac{g}{2 T}}+2 g^2 \psi ^{(1)}\left(\frac{g+T}{2 T}\right)+\frac{8 T e^{-k t} (-2 g k+g+T) \sinh (k t)}{k^2}\right),\label{var}
\end{eqnarray}
where $\mathcal{J}(t)$ is written in terms of the Gaussian hypergeometric functions 
\begin{eqnarray}
    \mathcal{J}(t)= \; _2F_1^{(0,2,0,0)}\left(\frac{g+T}{2 T},\frac{g+T}{2 T},\frac{g+T}{2 T},e^{-2 k t}\right)+ \\\nonumber  -2 \;_2F_1^{(1,1,0,0)}\left(\frac{g+T}{2 T},\frac{g+T}{2 T},\frac{g+T}{2 T},e^{-2 k t}\right)+\;_2F_1^{(2,0,0,0)}\left(\frac{g+T}{2 T},\frac{g+T}{2 T},\frac{g+T}{2 T},e^{-2 k t}\right),
\end{eqnarray}
with $_2F_1^{(n,m,l,0)}=\partial^3_{n,m,l}\left(_2F_1(n,m,l)\right)$, being the derivatives over the arguments of the Gaussian hypergeometric function. Because the mean is zero, the second moment is the variance, that is $\sigma_Q^2=\langle Q^2\rangle$. Apart from the complicated dependence in the parameters, the variance \ref{var} has a stationary behavior, which can be seen in  figure \ref{fig2} and derived analytically. For $t\rightarrow\infty$ the variance becomes
\begin{equation}
 \lim_{t\rightarrow\infty}   \sigma^2_Q = \frac{1}{4} \left(2 g^2 \psi ^{(1)}\left(\frac{g+T}{2 T}\right)+\frac{4 T (-2 g k+g+T)}{k^2}\right),
\end{equation}
where $\psi^{(1)}$ is the Poly-Gamma function \cite{abramowitz1988handbook}.

Moving forward, one can show that the third moment $\langle Q^3\rangle$ is also zero. This means that the skewness is null and suggests that the distribution could be symmetric around the mean (a behavior that can be qualitatively observed in Figure \ref{fig2} a), and numerically checked). As expected by checking the shape of the distribution in figure \ref{fig2} a). With the mean, variance, and skewness, it is only missing the kurtosis. The  excess  kurtosis for our case will simplified by the formula $$ \kappa_Q = \frac{\langle Q^4\rangle}{\langle Q^2\rangle^2 } -3.$$
This gives information about the shape of the distribution \cite{darlington_is_1970}. By  figure \ref{fig2} b) one can see that $\kappa_Q>-3$, which means that the distribution is leptokurtic \cite{darlington_is_1970}, meaning that the distribution in figure \ref{fig2} a) has fatter tails than the normal distribution. 
Besides, this expression could have been written analytically, due to Eq.~\ref{average}, but that we opt to omit here due to the extensive size of the expression. Nevertheless, the behavior is plotted in figure \ref{fig2}.

The graph in figure \ref{fig2} b) suggests that the even momenta of the heat reach stationary values. This leads us to the calculation of the asymptotic of the characteristic function $Z(\lambda)$, which will be given by
\begin{equation}
    \lim_{t\rightarrow\infty} Z(\lambda) = Z(\lambda,\infty) = \frac{(2k)^{\frac{g+T}{T}}}{\Gamma \left(\frac{g+T}{2 T}\right)^2}\phi(\lambda)\; _2F_1\left(z_\lambda,z_\lambda+ig\lambda;\frac{g+T}{2 T};0\right),
\end{equation}
where 
\begin{equation}
    \phi(\lambda) = \lim_{t\rightarrow\infty} f(\lambda)^{-z_\lambda} f({-\lambda})^{-z_\lambda-ig\lambda} \Gamma \left(z_\lambda\right) \Gamma \left(z_\lambda+ig\lambda\right),
\end{equation}
is finite and was used for shortening the notation. Due to the non-trivial dependence on $\lambda$, even in the asymptotic limit, the characteristic function cannot be integrated analytically to yield the expression of the heat distribution.

\section{Numerical Simulations}\label{sec5}

We can compare our results with numerical simulations, which give an indirect validation of the results. To generate the initial distribution of the initial position, we use the ``rejection method", where we used a uniform random variable to select the random points that obey the probability in Eq.~\ref{initial}. 
Using a fourth-order Runge-Kutta \cite{tocino_rungekutta_2002}, we integrate the SDE in  Eq.~\ref{langevin}, evolving the system up to $t=1$. The system studied has a singularity in $x=0$, which is not physically allowed, since we have an asymmetric trap, as pointed out in section \ref{sec2}. However, due to numerical errors, the simulation of the particle's position can become negative. One way to avoid this is to increase the precision of the numerical simulation. % Nevertheless, we can also ignore such results, using only the physically allowed.
With the ensemble of trajectories generated, we can measure the heat Eq.~\ref{heat}. Measuring different values of the heat, we can plot on a histogram and compare with the theoretical result. 

The histogram and the theoretical result are plotted in figure \ref{fig1}, fitting the  exact curve quite well,  therefore indirectly validating our results via the numerical simulations.

\section{Discussion and Conclusion}\label{sec6}

In the present paper, we have studied a diffusive system in a logarithm-harmonic potential. This system presents a simple stochastic energetic characterization: it is described only by heat and internal energy. Herein, we derived the heat distribution exactly. The result is compared with numerical simulations and it is in good accordance. Moreover, we derive analytically the characteristic function for the heat probability distribution, allowing us to calculate all the moments of the distribution, and to characterize all the statistical behavior of the system. 

The heat distribution is symmetric in $Q$, which means that $P(Q)=P(-Q)$ in agreement with the vanishing skewness and figure \ref{fig2} a).  We must notice that this is a consequence of choosing an equilibrated initial distribution. This equilibrium, non entropy producing dynamics, implies that the particle has the same chance of absorbing or losing energy from the heat bath. This symmetric behavior can be understood by noticing that the particle can equilibrate with the heat bath, a behavior which is not present in the case without the harmonic contribution \cite{paraguassu_heat_2021}. Moreover, if we compare with the single harmonic case \cite{chatterjee_exact_2010}, one can see that our distribution has a similar shape to the heat distribution in \cite{chatterjee_exact_2010}, but  in the presence of the logarithm potential the divergence in $Q=0$ is not present. Moreover, besides the apparent fast decay of the distribution, one can show that the tails of the distribution decay slower than an exponential decay, forbidding the usual analysis via large deviation theory \cite{touchette_large_2009}.

The fluctuations of the heat are described by the characteristic function $Z(\lambda)$. In our case, $Z(\lambda)$ has a non-trivial dependence due to the Gaussian hypergeometric function, together with the Gamma functions. That not withstanding, by means of  $Z(\lambda)$, we calculate the mean, variance, skewness, and kurtosis. The mean, skewness, and any odd moment are zero, as expected since the probability is symmetric in $Q$. Moreover, as plotted in figure \ref{fig2} b), the evolution of the variance and the kurtosis reach a stationary value, which can be understood as a consequence of the particle starting comming to equilibrium with the heat bath. We calculate the stationary behavior for the variance and also for the characteristic function. Writing in terms of Poly-Gamma functions, Gamma functions, and Gaussian hypergeometric, the asymptotic characteristic function is still not integrable in $\lambda$. Reinforcing the non-trivial nature of the statistical behavior of the heat.

In conclusion, we have obtained exact results for the heat distribution, the characteristic function, and its moments. Our results are compared with numerical simulations, showing its consistency. The results are discussed and compared with the literature, giving new insights into the problem. Moreover, our finding extends the list of characterized heat distribution in stochastic thermodynamics.

\section*{Acknowledgments}
We would like to thanks Victor Valad\~ao and Lucianno Defaveri for useful discussions. This work is supported by the Brazilian agencies CAPES and CNPq. P.V.P. would like to thank CNPq for his current fellowship. This study was financed in part by Coordena\c c\~ ao de Aperfei\c coamento de Pessoal de N\' ivel Superior - Brasil (CAPES) - Finance Code 001.

\section*{References}

\bibliography{name.bib}
\bibliographystyle{name.bst}

\end{document}